\documentclass[preprint2, times, twocolappendix]{aastex631}

\newcommand{\projecttitle}{Gnuastro: measuring radial profiles from images}
\newcommand{\projectversion}{104aad5}
\newcommand{\projectgitrepo}{https://codeberg.org/gnuastro/papers}
\newcommand{\projectgitbranch}{radial-profile}
\newcommand{\projectcopyrightowner}{Raul Infante-Sainz <infantesainz@gmail.com>}
\newcommand{\gnuastroversion}{0.21.43-3101}

\newcommand{\maneageversion}{8161194}

\newcommand{\chr}{iSDSS}
\newcommand{\chg}{rSDSS}
\newcommand{\chb}{gSDSS}

\newcommand{\chrefband}{rSDSS}

\newcommand{\jplusdr}{3}

\newcommand{\profsclipmultiple}{3}

\newcommand{\figxmintop}{-0.462999}
\newcommand{\figxmaxtop}{15.742}
\newcommand{\figxminbottom}{-50}
\newcommand{\figxmaxbottom}{1700}

\ifdefined\highlightnew

\else

\fi

\ifdefined\highlightnotes
\newcommand{\tonote}[1]{\textcolor{red!60!black}{[#1]}}
\else
\newcommand{\tonote}[1]{{}}
\fi

\usepackage{graphicx}

\usepackage{xcolor}
\color{black} % Color of main text.
\definecolor{DarkBlue}{RGB}{0,0,90}

\hypersetup{
    pdftitle={\projecttitle},
    pdfauthor={\projectcopyrightowner},
    pdfsubject={\projectgitrepo{} (commit \projectversion)},
    pdfkeywords={Reproducible research, Maneage, ADD YOUR OWN}
}

\usepackage{tikz}
\usetikzlibrary{external}
\tikzsetexternalprefix{tex/tikz/}

\usepackage{pgfplots}
\pgfplotsset{compat=newest}
\usepgfplotslibrary{groupplots}
\usetikzlibrary{plotmarks}
\pgfplotsset{
  axis line style={thick},
  tick style={semithick},
  tick label style = {font=\footnotesize},
  every axis label = {font=\footnotesize},
  legend style = {font=\footnotesize},
  label style = {font=\footnotesize}
  }

\shorttitle{\projecttitle}
\shortauthors{Infante-Sainz, R.}

\begin{document}

%% Title
\title{\projecttitle}

%% Authors
\author[0000-0002-6220-7133]{Ra\'ul Infante-Sainz}
\author[0000-0003-1710-6613]{Mohammad Akhlaghi}
\author[0000-0002-6672-1199]{Sepideh Eskandarlou}
\affiliation{Centro de Estudios de F\'isica del Cosmos de Arag\'on (CEFCA), Plaza San Juan, 1, E-44001, Teruel, Spain}

\correspondingauthor{Ra\'ul Infante-Sainz}
\email{infantesainz@gmail.com}

%% Abstract
\begin{abstract}
  \noindent
  Radial profiles play a crucial role in the analysis and interpretation of astronomical data, facilitating the extraction of spatial information.
  However, highly customizable (for different scenarios) measurements over each elliptical annulus can be challenging.
  In response, we present \texttt{\small{astscript-radial-profile}}, which is part of Gnuastro from version~0.15 and has an extensive documentation.
  A convenient feature of this program is its capability to make the measurements with different operators (mean, median, sigma-clipping, and many more) over ellipses, very quickly and directly on the command-line with minimal dependencies.
  This research note is reproducible with Maneage, on the Git commit \projectversion.
\end{abstract}

%% Keywords (from https://astrothesaurus.org)

%% Start of main body.
\section{Introduction}
\label{sec:introduction}
\noindent
A radial profile consists of measurements over elliptical annuli at increasing distances/radii.
In general, a radial profile can be generalized to any number of dimensions.
For example, in a 3D dataset (cube), a radial profile consists of measurements over growing ellipsoidal shells.
Radial profiles are most commonly employed in 2D images because the third dimension in 3D astronomical data (IFUs or radio data cubes) is spectral, not spatial.
The radial profile is a table with two columns: the distance to the center and the measured quantity.

Even with good visualization tools and proper color bars set, quantifying the spatial variation of an object's brightness, and comparing it with other objects is challenging.
A radial profile is therefore a convenient way of extracting spatial information of an astronomical target.

Astronomical sources can have irregular shapes (for example stellar streams and Galactic cirrus).
In these cases, it is possible to compute radial profiles for a specified azimuth angle or direction.
More generally, astronomical sources, driven by gravitational forces, tend to exhibit symmetrical, often ellipse-symmetric shapes.
As a consequence, this type of aperture is very useful to derive physical properties such as surface brightness profiles, mass density profiles, and star formation rate profiles, among others, as demonstrated in recent studies \citep[see e.g., ][]{infantesainz2020,liu2022,monsekov2020,martinezlombilla2023,watkins2023}.

\begin{figure*}
  \ifdefined\makepdf%
    \tikzsetnextfilename{fig-m81-profile}%
    \input{tex/src/fig-m81-profile.tex}%
  \else
    \includegraphics[width=\linewidth]{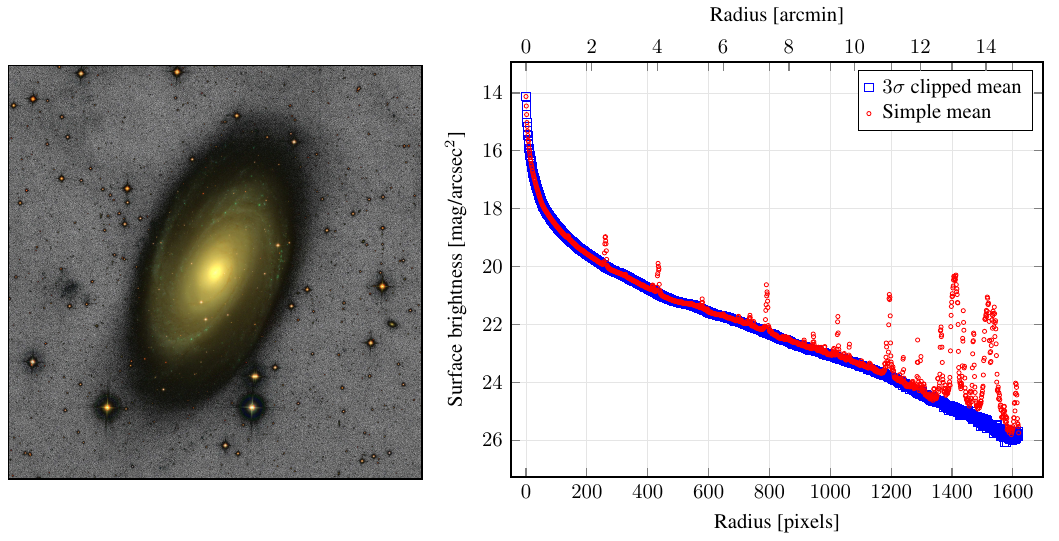}
  \fi

  \vspace{-7mm}
  \caption{
   \label{fig:m81}
   Color image and surface brightness radial profiles of M81 galaxy.
   \emph{Left:} Color image with gray background obtained by using the \texttt{\small{astscript-color-faint-gray}} script \citep{infantesainz2024}.
   \emph{Right:} Surface brightness radial profiles of the \chrefband{} image, measured with two different operators.
   The y-axis represents the surface brightness (SB), while the x-axis represents the distance to the center of the galaxy.
   The red circles correspond to the measurements obtained by a simple mean, and they exhibit peaks due to the bright objects (stars).
   In contrast, the blue squares represent the measurements obtained using a $\profsclipmultiple\sigma$ clipped mean, resulting in a smoother profile by removing the extreme outlier values (for example those from contaminant sources).
}
\end{figure*}

Once the annuli are defined, selecting the appropriate operator for measuring the properties becomes crucial.
While simple mean or sum measurements are common, other situations require more sophisticated operators.
For instance, measuring the brightness within each radius with a simple mean or sum can be severely affected by outliers at certain radii: foreground or background objects or star-forming regions within the galaxy itself (see Figure~\ref{fig:m81}).
As a consequence, robust estimators of the measurement such as median or sigma clipping operators are desired, but a diverse set of measurements are rarely implemented in many implementations of radial profiles.

In this research note, we present a powerful solution to address these problems: the Gnuastro executable program \texttt{\small{astscript-radial-profile}}.
This program addresses the complexities arising from the generation of radial profiles on astronomical images.
It is able to create radial profiles of astronomical sources very fast, from the command-line, and with an enormous variety of measurements\footnote{Measurements: \url{https://www.gnu.org/software/gnuastro//manual/html_node/MakeCatalog-measurements.html}}.
The following section provides an overview, using the radial profile of the M81 galaxy as an illustrative example.

\section{Creating radial profiles}
\label{sec:creatingradialprofiles}
\noindent
Gnuastro's \texttt{\small{astscript-radial-profile}} is an installed script (written in POSIX shell), and it uses several Gnuastro programs (that are installed with it and available on the host operating system) to generate the radial profile.
It is possible to run the script and keep the intermediate files (passed between the programs) to inspect and optimize the execution for each particular science case.
A summary of the general steps are described below.

\begin{itemize}
  \item Aperture construction.
        The script uses Gnuastro's MakeProfiles (\texttt{\small{astmkprof}}) to construct elliptical annuli based on user-defined parameters, including the center position, axis ratio, position angle, azimuthal angle, and others.
  \item Measurement extraction.
        Gnuastro's MakeCatalog (\texttt{\small{astmkcatalog}}) is then used to obtain the requested measurements over the elliptical annuli generated previously.
  \item Output generation.
        The script produces a formatted table (possibly FITS) containing the resulting radial profile.
\end{itemize}

By default, if no options are specified, the program computes the radial profile using circular apertures centered on the input image center, employing a simple mean as the measurement operator.
To customize the radial profile, users can modify multiple options and parameters, allowing adjustments to the ellipse positioning, shape, and measurement operators.
Beyond the large number of possible measurements, we find particularly interesting the possibility of changing the size of the apertures by undersampling or oversampling, as well as the generation of radial profiles with sub-pixel precision.
For a comprehensive understanding of the different parameters and how to use this script in more detail, we refer to the Gnuastro manual\footnote{Manual: \url{https://www.gnu.org/software/gnuastro/manual}}.

The left panel of Figure~\ref{fig:m81} depicts a color (and gray background) visualization of the M81 galaxy captured from the J-PLUS DR\jplusdr{} \citep{cenarro2019} using the \chr, \chg, and \chb{} filters.
This color image has been obtained with another Gnuastro's installed scripts \texttt{\small{astscript-color-faint-gray}} script \citep[for more details, see][]{infantesainz2024}.
In this image there are several other sources and over M81, star-forming regions produce non elliptically-symmetric features.
The right panel of Figure~\ref{fig:m81} shows the radial profiles computed by using two different operators: simple mean and sigma-clipped mean.
The radial profiles look quite different.
While the simple mean radial profile has several peaks that are due to the presence of nearby objects (bright stars in the surroundings of the M81 galaxy), the sigma clipping radial profile is smoother and does not have these peaks.

\section{Acknowledgments}
\label{sec:acknowledgments}
\noindent
The workflow uses Maneage \citep[\emph{Man}aging data lin\emph{eage},][commit \maneageversion]{maneage}.
This research note is created from the Git commit {\projectversion}, hosted on Codeberg\footnote{Git repository of the paper (\texttt{\small{\projectgitbranch}} branch): \url{\projectgitrepo}.} which is archived on Software Heritage\footnote{Software Heritage (SoftWare Hash IDentifier, SWHID): \href{https://archive.softwareheritage.org/swh:1:dir:d5029e066916cb64f0d95d20eb88294acc78b2b1;origin=https://codeberg.org/gnuastro/papers;visit=swh:1:snp:b065324c2ef3b48bc26e8f30e48102a1abd2052f;anchor=swh:1:rev:61764447b16da44538e5ddbf7fb69937ba138e81}{swh:1:dir:d5029e066916cb64f0d95d20eb88294acc78b2b1}} for longevity.
Supplements are also available on Zenodo\footnote{Zenodo: \url{https://doi.org/10.5281/zenodo.10124582}}.

The analysis uses GNU Astronomy Utilities \citep[Gnuastro,][]{gnuastro2015,gnuastro2019} v\gnuastroversion.
Gnuastro has been funded by the Japanese MEXT scholarship and its Grant-in-Aid for Scientific Research (21244012, 24253003), ERC 339659-MUSICOS, Spanish MINECO AYA2016-76219-P, and NextGenerationEU ICTS-MRR-2021-03-CEFCA.
We acknowledge the funding by Governments of Spain and Arag\'on through FITE and Science Ministry (PGC2018-097585-B-C21, PID2021-124918NA-C43).

%% Bibliography
\bibliography{references}{}
\bibliographystyle{aasjournal}

%% Appendix. Mention all used software.
\appendix
\section{Software acknowledgement}
\label{appendix:software}
 
This research was done with the following free software programs and libraries:  1.23, Bzip2 1.0.8, C compiler (Apple clang version 13.1.6 (clang-1316.0.21.2.5)), CFITSIO 4.1.0, CMake 3.24.0, Dash 0.5.11-057cd65, Discoteq flock 0.4.0, Expat 2.4.1, File 5.42, Fontconfig 2.14.0, FreeType 2.11.0, GNU AWK 5.1.1, GNU Astronomy Utilities 0.21.43-3101 \citep{gnuastro2015,gnuastro2019}, GNU Autoconf 2.71, GNU Automake 1.16.5, GNU Bash 5.2-rc2, GNU Bison 3.8.2, GNU Coreutils 9.1, GNU Diffutils 3.8, GNU Findutils 4.9.0, GNU Grep 3.7, GNU Gzip 1.12, GNU Libtool 2.4.7, GNU M4 1.4.19, GNU Make 4.3, GNU Multiple Precision Arithmetic Library 6.2.1, GNU Multiple Precision Floating-Point Reliably 4.1.0, GNU NCURSES 6.3, GNU Nano 6.4, GNU Readline 8.2-rc2, GNU Scientific Library 2.7, GNU Sed 4.8, GNU Tar 1.34, GNU Texinfo 6.8, GNU Wget 1.21.2, GNU Which 2.21, GNU gettext 0.21, GNU gperf 3.1, GNU libiconv 1.17, GNU libunistring 1.0, GPL Ghostscript 9.56.1, Git 2.37.1, Help2man , Less 590, Libffi 3.4.2, Libgit2 1.3.0, Libidn 1.38, Libjpeg 9e, Libpaper 1.1.28, Libpng 1.6.37, Libtiff 4.4.0, Libxml2 2.9.12, Lzip 1.23, OpenSSL 3.0.5, Perl 5.36.0, Python 3.10.6, WCSLIB 7.11, X11 library 1.8, XCB-proto (Xorg) 1.15, XZ Utils 5.2.5, Zlib 1.2.11, cURL 7.84.0, libICE 1.0.10, libSM 1.2.3, libXau (Xorg) 1.0.9, libXdmcp (Xorg) 1.1.3, libXext 1.3.4, libXt 1.2.1, libpthread-stubs (Xorg) 0.4, libxcb (Xorg) 1.15, pkg-config 0.29.2, podlators 4.14, util-Linux 2.38.1, util-macros (Xorg) 1.19.3, xorgproto 2022.1 and xtrans (Xorg) 1.4.0. 
The \LaTeX{} source of the paper was compiled to make the PDF using the following packages: courier 61719 (revision), epsf 2.7.4, etoolbox 2.5k, helvetic 61719 (revision), lineno 5.3, pgf 3.1.10, pgfplots 1.18.1, revtex4-1 4.1s, tex 3.141592653, textcase 1.04 and ulem 53365 (revision). 
We are very grateful to all their creators for freely  providing this necessary infrastructure. This research  (and many other projects) would not be possible without  them.

%% Finish LaTeX
\end{document}